\documentclass[twocolumn,aps,prl,showpacs,preprintnumbers,amsmath,amssymb,superscriptaddress,10pt]{revtex4}
\pdfoutput=1
\usepackage{graphicx}% Include figure files
\usepackage{dcolumn}% Align table columns on decimal point
\usepackage{bm}% bold math
\usepackage{dcolumn}
\usepackage{braket}
\usepackage{amsmath}
\usepackage{array}
\usepackage{color}
\begin{document}

\title{Skew scattering in dilute ferromagnetic alloys}

\author{Bernd Zimmermann}
\affiliation{Peter Gr{\"u}nberg Institut and
Institute for Advanced Simulation, Forschungszentrum J{\"u}lich and JARA, Germany}
\author{Kristina Chadova}
\affiliation{Department of Chemistry, Physical Chemistry,
Ludwig-Maximilians University Munich, Germany}
\author{Diemo K\"odderitzsch}
\affiliation{Department of Chemistry, Physical Chemistry,
Ludwig-Maximilians University Munich, Germany}
\author{Stefan Bl{\"u}gel}
\affiliation{Peter Gr{\"u}nberg Institut and
Institute for Advanced Simulation, Forschungszentrum J{\"u}lich and JARA, Germany}
\author{Hubert Ebert}
\affiliation{Department of Chemistry, Physical Chemistry,
Ludwig-Maximilians University Munich, Germany}
\author{Dmitry V. Fedorov}
\affiliation{Max Planck Institute of Microstructure Physics, Weinberg 2,
06120 Halle, Germany}
\affiliation{Institute of Physics, Martin Luther University Halle-Wittenberg,
06099 Halle, Germany}
\author{Nguyen H. Long}
\affiliation{Peter Gr{\"u}nberg Institut and
Institute for Advanced Simulation, Forschungszentrum J{\"u}lich and JARA, Germany}
\author{Phivos Mavropoulos}
\affiliation{Peter Gr{\"u}nberg Institut and
Institute for Advanced Simulation, Forschungszentrum J{\"u}lich and JARA, Germany}
\author{Ingrid Mertig}
\affiliation{Max Planck Institute of Microstructure Physics, Weinberg 2,
06120 Halle, Germany}
\affiliation{Institute of Physics, Martin Luther University Halle-Wittenberg,
06099 Halle, Germany}
\author{Yuriy Mokrousov}
\email{y.mokrousov@fz-juelich.de}
\affiliation{Peter Gr{\"u}nberg Institut and
Institute for Advanced Simulation, Forschungszentrum J{\"u}lich and JARA, Germany}
\author{Martin Gradhand}
\email{m.gradhand@bristol.ac.uk}
\affiliation{H.~H.~Wills Physics Laboratory,
University of Bristol, Bristol BS8 1TL, United Kingdom}

\date{\today}

\begin{abstract}
The challenging problem of skew scattering for  Hall effects in dilute ferromagnetic alloys, with
intertwined effects of spin-orbit coupling, magnetism and impurity scattering, is studied here 
from first principles.  Our main aim is to identify chemical trends and work out simple rules
for large skew scattering  in terms of the impurity and host states at  the Fermi surface, with 
particular emphasis on the interplay of the spin and anomalous Hall effects in one and the same system.  
The predicted trends are  benchmarked 
by referring to three different \emph{ab initio} methods based on different approximations 
with respect to the electronic structure and transport properties.
\end{abstract}

\pacs{71.15.Rf, 72.25.Ba, 75.47.Np, 85.75.-d}

\maketitle

The anomalous Hall effect (AHE) was discovered in 1881~\cite{Hall1881} but kept its secrets 
for a very long time. It took more than 70 years to establish the spin-orbit coupling (SOC) as the driving force behind the phenomenon~\cite{KL_54, Smit_55, Nagaosa2010}.
Since that time, the main stream of research was directed at identifying  and understanding the various microscopic
mechanisms~\cite{Nagaosa2010, Smit_58, MottMass65, Landau65, Berger_1970, Dyakperel71, Berger_1972, Crepieux01, Sinova_04, Sinitsyn_rev08}
contributing to the total effect as observed in experiment. This work was driven by experimental~\cite{McAlister1976, McAlister1977, Hurd1977, Hurd1979, Fert1980, Fert1981, Matveyev1982, Xiong1992, Miyasato2007, Tian2009} as well as theoretical~\cite{Sinitsyn2005, sinitsyn_06, Onoda2006, Sinitsyn_rev08, Lowitzer2010a, Weischenberg2011, Czaja2014, Fert2011} progress in the decoding of the microscopic processes leading
to the AHE. The established separation is along the lines of intrinsic bandstructure induced effects~\cite{KL_54,Thouless1982, Xiao2010} and extrinsic contributions related to scattering at perturbations~\cite{Smit_58, MottMass65, Landau65, Berger_1970, Dyakperel71, Berger_1972}. The dominance of specific mechanisms under different conditions has been under debate for decades but was recently settled on a general basis~\cite{Onoda2006, Sinitsyn_rev08}.

Importantly, the underlying principles of the AHE are equivalent to those responsible for the spin Hall effect (SHE). Since 
it was realized that
the SHE has the potential to drastically change the way spin currents are created in spintronic devices the AHE experienced a revival. Phenomenologically, the only difference between the two effects is the ferromagnetic order needed for the AHE, while the SHE 
exists also in nonmagnetic materials. In terms of practicality, the existence of a finite Hall voltage makes the AHE much easier accessible than the SHE which creates a spin imbalance only. 
Ultimately, the SHE and the AHE are the archetypical transport phenomena for the exploration of spin-orbit
coupling where the motion of charge carriers creates transversal spin currents and vice versa. Their understanding will pave the way to related thermoelectric phenomena such as the anomalous and spin Nernst effects~\cite{Bergman2010, Tauber2012, Tauber2013, Weischenberg2013, Wimmer2013, Wimmer2014}.

Among various contributions to the AHE and SHE of intrinsic and extrinsic origin, the 
{\it skew scattering} provides the dominant source of transverse current in the limit of dilute 
alloys~\cite{Onoda2006, Sinitsyn_rev08}. 
The reason is the linear scaling of the skew-scattering driven transverse conductivity
$\sigma_{yx}$ with the diagonal conductivity $\sigma_{xx}$ for vanishing scattering. 
The corresponding scaling
constants, the so-called anomalous or spin Hall angles, AHA or SHA, are respectively 
defined as
\begin{equation}
\alpha_{\text{AHE}}=\sigma_{yx}/\sigma_{xx}, \quad \alpha_{\text{SHE}}=\sigma^s_{yx}/\sigma_{xx},
\end{equation}
where superscript $s$ refers to the spin conductivity tensor. While it is far from trivial
to access the Hall angles in experiment directly, they play a pivotal role in 
spintronic studies which hinge on transverse current generation via Hall effects.
It is well-known that the value of the Hall angle derived in experiment will strongly depend on the material composition and preparation~\cite{Miyasato2007,Tian2009}. It is thus of crucial importance to 
achieve material-specific theoretical understanding of the skew scattering for the purposes of engineering the
desired functionalities of spintronic devices.

To this end, the first-principles assessment of skew scattering for the case of the spin Hall effect has been implemented for paramagnets only~\cite{Gradhand2010b,Lowitzer2011}. 
In case of ferromagnets, however, the situation is far more complex owing to the subtle interplay of the magnetization with spin-orbit interaction. Moreover, the magnetism in transition-metals is normally
carried by localized $d$ and $f$ electrons whose presence at the Fermi energy, $E_F$, results in complex multi-sheeted Fermi surfaces. This
prohibits the analysis in terms of simple models for scattering, such as,~e.g., the phase shift 
model~\cite{Fert2011,Johansson2014}. Nevertheless, experimentally the SHE in ferromagnets has been discussed recently \cite{Wei2012,Miao2013}, where for the case of Ref.~\cite{Wei2012} the underlying mechanism is most likely the skew scattering in Pd(Ni) dilute alloys.

In this Letter, we explore both the AHE and SHE in dilute {\it ferromagnetic} alloys. Using first-principles methods, we provide insights into the physics of the skew-scattering mechanism, which governs the considered phenomena in the dilute limit. For both, magnetic and nonmagnetic hosts, we analyse chemical trends to draw general conclusions. The vast range of values in the Hall
angle, which we present, provides an opportunity to engineer materials according to specific requirements.
So far the skew scattering for SHE in ferromagnets has not been studied from first
principles, and here we demonstrate that in the considered alloys it can be rather prominent. At the same time, for given host impurity combination the SHE and AHE are intrinsically correlated  showing similar overall trends and sign changes. 

\begin{table}[t]
  \caption{Three different first-principles approaches used for the calculations presented in this Letter. The abbreviations stand for: BE $-$ Boltzmann equation, KSF $-$ Kubo-St\v{r}eda formula, FP $-$ full potential, and ASA $-$ atomic sphere approximation.}
\begin{ruledtabular}
\begin{tabular}{c||c|c|c}
Approach       &   Transport    &  Electronic & Spin-orbit \\
 &   description    &  structure & coupling\\
\hline
\rule{0pt}{3ex}
Method A       &   BE       &  FP &  Pauli  \\
%\hline
Method B       &   BE       &  ASA & Dirac \\
%\hline
Method C       &  KSF   & ASA & Dirac \\
\end{tabular}
\end{ruledtabular}
\end{table}

Owing to the complexity of the problem outlined above, 
we have chosen to compare and benchmark three distinct state-of-the-art first-principles approaches to arrive at sound conclusions. As we shall see, many degrees of freedom
in relativistic ferromangetic transition-metal systems can influence the results significantly, which makes the material-specific predictions for the Hall angle very delicate and sensitive to the details of the 
electronic structure.

The three methods used are briefly introduced below and summarized in Table~I.
As for the SOC, it is included within the Pauli approach in method A, while methods B and C are based on the solution of the fully-relativistic Dirac equation.
Methods B and C rely on the atomic sphere approximation (ASA) in contrast to the full potential (FP) description of method A. For computing the transport properties methods A and 
B exploit the semiclassical picture in terms of the Boltzmann equation (BE)~\cite{Gradhand2010b}. Considering cubic crystals and sign conventions from Ref.~\cite{Fedorov2013}, the 
$yx$ component of the conductivity tensor (anomalous Hall conductivity, AHC) is computed as
\begin{equation}\label{Eq.:1}
  \sigma_{yx}= \frac{{\rm e}^2}{\hbar} \, \frac{1}{\left( 2\pi \right)^3} ~ \int_{\rm FS} {\rm d}S ~ { \frac{ v_{y}({\bf k}) \, \lambda_{x}({\bf k}) }{\left\lvert {\bf v}({\bf k}) \right\rvert }  }\ \text{,}
\end{equation}
where FS stands for the Fermi surface integration, $\mathbf{v} (\mathbf{k})$ and $\boldsymbol\lambda (\mathbf{k})$  are the group velocity and the mean free path, respectively. The latter is determined as the self-consistent solution of the integral Boltzmann equation which takes as input the scattering matrix at a given isolated impurity in a particular host. The spin Hall conductivity 
$\sigma_{yx}^s$ (SHC) is computed similarly, taking into account the spin polarization of electron states~\cite{Gradhand2010b}. Method C 
employs
 the Kubo-St\v{r}eda formula in combination with the coherent-potential approximation including
chemical
 disorder to compute the conductivities~\cite{Lowitzer2010a,Lowitzer2011,Streda_82}. In this approach all contributions to the Hall effect are treated on equal
footing and the Hall angles are determined from Eq.~(1) in the limit of vanishing impurity concentration.

\begin{figure}[t!]
\includegraphics[width=0.9\linewidth]{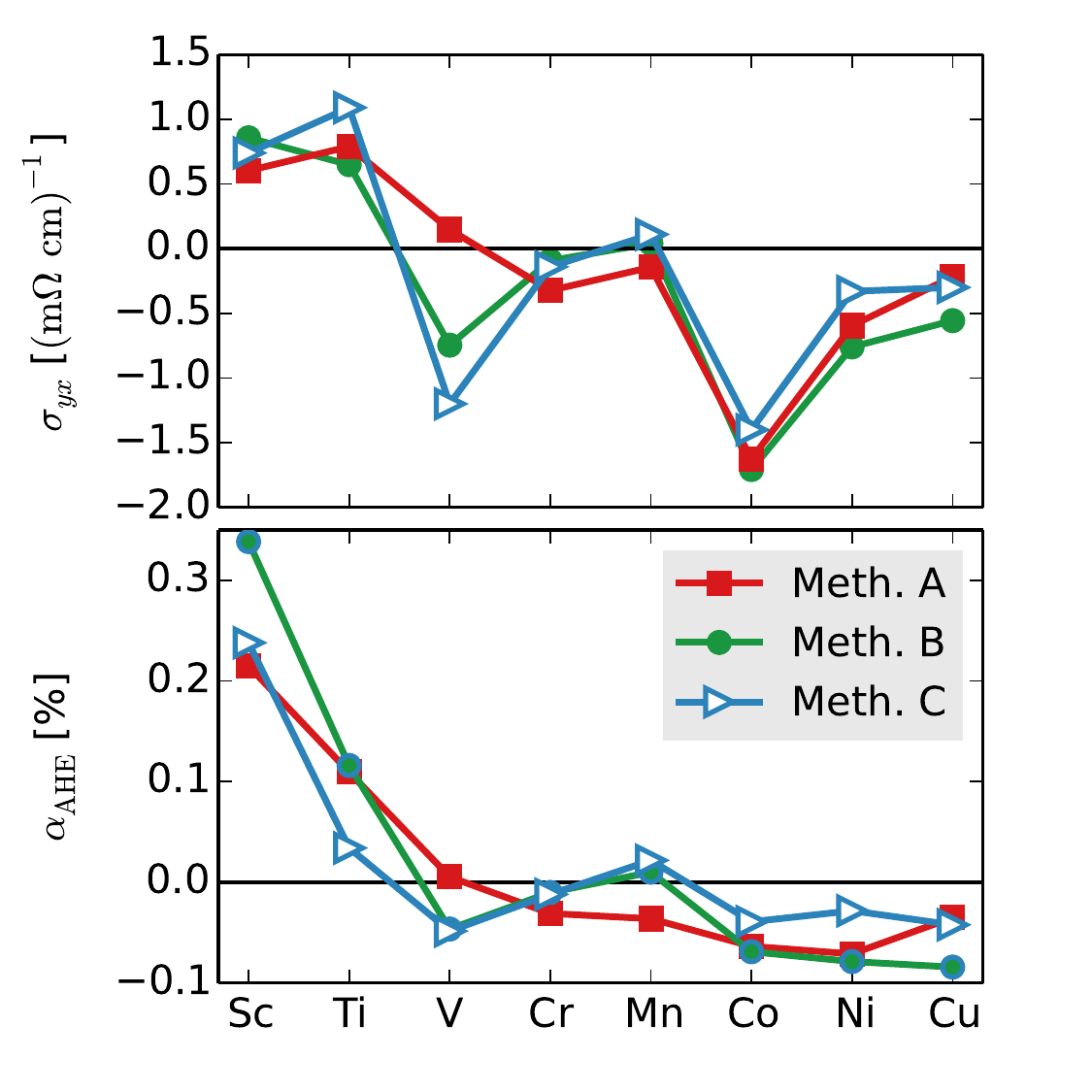}
\caption{Skew-scattering contribution to the anomalous Hall conductivity (upper panel) and the anomalous
Hall angle (lower panel) for the ferromagnetic Fe host with $3d$  impurities with concentration of 1~at.~\%.
 \label{fig:Fe2}}
\end{figure}

As a first example we consider the prototype ferromagnet $-$ bcc Fe $-$ doped with $3d$ impurities from Sc to Cu. As evident from Fig.~\ref{fig:Fe2}, where the results of the calculations
for the AHA and $\sigma_{yx}$ with all three methods are presented, all approaches  agree not only in magnitude, but also in the general trend of the AHA and AHC with a characteristic change of sign as the impurity is varied along the 3$d$ series. 
Noticeably, our results show that the acquired AHA does not exceed a tiny value of 0.1\%, with 
the only exception of Fe(Sc).
Furthermore, there is a difference in magnitude and even sign for V and Mn
impurities, which 
we were able to trace back to the difference in the FP and ASA description of the electronic structure
of Fe, with slightly different relative position of the Fe $d$-states with respect to the Fermi level. 
Overall, we underline that
the magnitude and sign of $\sigma_{yx}$ (AHC) depends drastically on the host-impurity combination and on the details of charge density screening around the impurity, 
which motivates the use of {\it ab-initio} methods for understanding the physics creating the skew scattering in transition metals~\cite{Suppl1}.

Based on these results and our previous experience with paramagnetic systems~\cite{Gradhand_10_d,Gradhand2010b}, we formulate 
the universal condition
 for strong skew scattering in spin-polarized situation:  the effective SOC, defined as the difference between the SOC strength of the host and the impurity, has to be large. Based on this criterion the small magnitude of the
AHA in the Fe-based systems from before can be explained by the very small effective 
SOC. Thus, from the point of view of the SOC strength, in order for a material to have strong 
skew scattering, the presence of  heavy transition metals is necessary. One route to achieve this has been intensively explored in the past experimentally for the AHE~\cite{Hurd1977, Hurd1979,Fert1980, Fert1981, Matveyev1982, Xiong1992, Miyasato2007} and it lies in a combination of a heavy metal host with $3d$ magnetic impurities.  In the remainder of this work, we choose Pd, Pt, and Au as examples for working out a microscopic condition for strong skew scattering not only for the AHE, but simultaneously for the SHE, also present and partially experimentally explored in these materials~\cite{Wei2012, Miao2013}.

\begin{figure}[t!]
\includegraphics[width=0.9\linewidth]{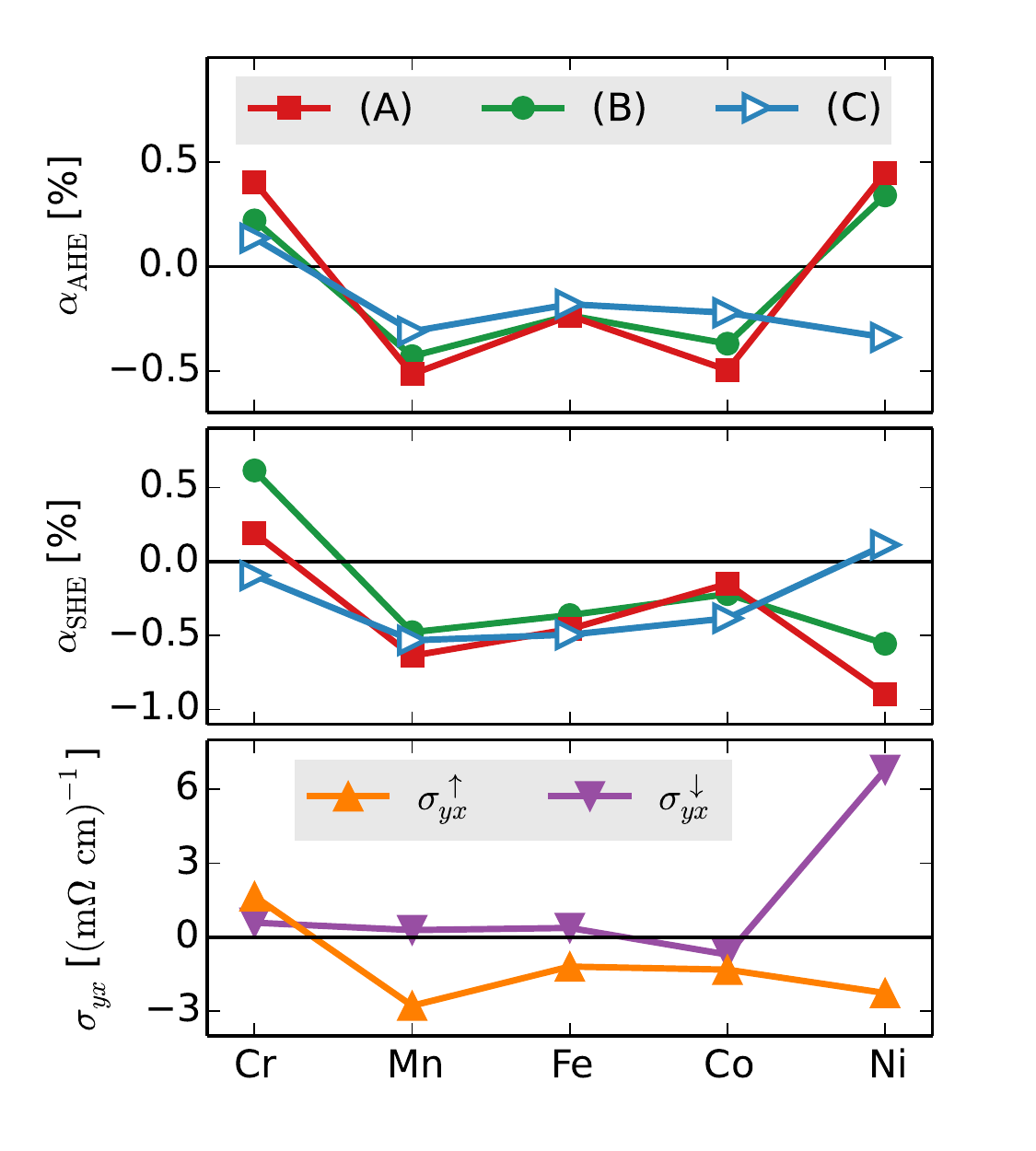}
\caption{
Computed with three different methods skew-scattering contribution to the AHA (upper panel), the SHA (middle panel), together with
spin-resolved conductivities (lower panel, method A only) 
in five dilute alloys based on a Pt host with an impurity concentration of 1~at.~\%. 
\label{fig:Pt}}
\end{figure}

\begin{figure}[t!]
\includegraphics[width=0.9\linewidth]{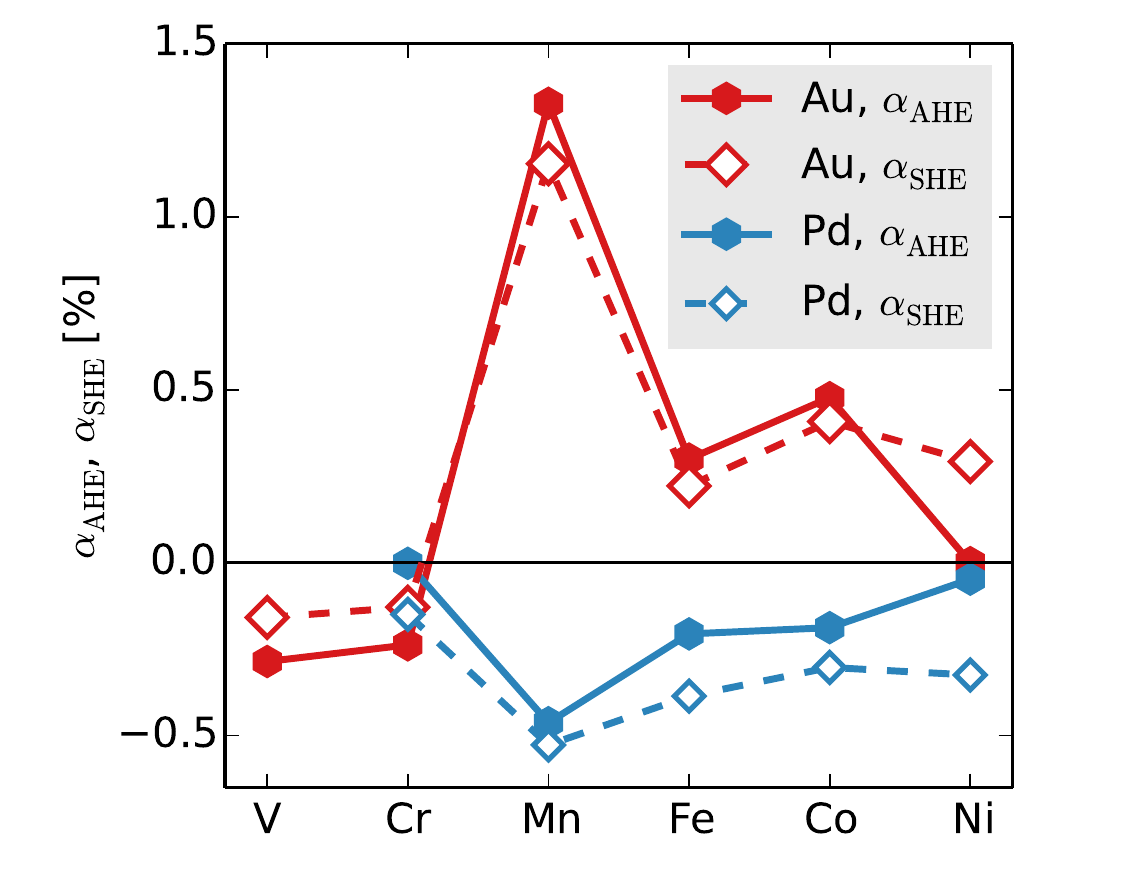}
\caption{
Skew-scattering contribution to the AHA and SHA 
in alloys based on the non-magnetic Au and Pd hosts with the magnetic $3d$ impurities (concentration of 1~at.~\%) as computed with method A. The non-magnetic Au(Ni) system is presented for reference. 
\label{fig:Au_Pd}}
\end{figure}

We first turn to Pt host considering all magnetic $3d$ impurities assuming ferromagnetic order with the magnetization along the $[00\bar{1}]$-direction. This corresponds to a typical AHE measurement in an applied external magnetic field. Our results obtained with all three approaches for the AHA and the SHA are shown in Fig.~\ref{fig:Pt}.
One immediately notices the large magnitude of the obtained Hall angles, which are almost an order of magnitude 
larger than in the respective Fe dilute alloys. Remarkably, the magnitude of the SHA is comparable to that
of the AHA in these systems. Moreover, with the only exception of Ni impurities within all three approaches and Cr impurities
as computed with method C, the sign of the AHA and SHA is in one-to-one correspondence. As shown in Fig.~\ref{fig:Au_Pd} we also observe a similar trend for the Au and Pd hosts with the magnetic $3d$ impurities from V (Cr) to Co (Ni). 

To understand the obvious correlation between the SHA and AHA, we analyse the spin-resolved Hall conductivities defined as $\sigma_{yx}^\uparrow=(1/2)(\sigma_{yx}+\sigma_{yx}^{s})$ and     $\sigma_{yx}^\downarrow=(1/2)(\sigma_{yx}-\sigma_{yx}^{s})$, which, within the two-current model, would correspond 
to the conductivities of the spin-up and spin-down electrons, respectively. The values of the spin-resolved conductivities computed with method A for Pt and presented in Fig.~\ref{fig:Pt},
point at consistent suppression of the skew-scattering for spin-down electrons in Pt doped with Cr, Mn,
Fe and Co. The situation in Au and Pd (not shown)
is exactly analogous to that in Pt. Thus, in the majority of considered systems 
the transverse current which is responsible for both AHE and SHE is almost purely spin-up polarized
(in Pt(Ni) the situation is reversed). 

The reason behind this can be explained based on the local densities of states (LDOS) of the host and of the impurity atoms.
Taking the Pt host, for which the DOS is dominated by the $d$-electrons at $E_F$, and Mn impurity as a
representative defect,
we can understand the weak spin-up scattering with enhanced $\sigma_{yx}^\uparrow$ from the fact
of the similar behaviour and orbital character of the host and impurity LDOS at $E_F$: the spin-up Mn LDOS at $E_F$  is also predominantly of $d$-character. 
For the spin-down Mn LDOS the $d$-resonance is pushed to higher energies due to the exchange splitting, leading to a more prominent $s$-like orbital character at $E_F$ - hence the host and the impurity LDOS are different,
and the scattering for spin-down electrons is stronger. 
The same behaviour exists for the spin-split conductivities
in Au. This can be explained from the free-electron-like character of the states at $E_F$ in Au, while the Mn impurity states share this character for spin-up states, a $d$-resonance is present for the spin-down channel at $E_F$. The analysis can be extended to Cr, Fe, and Co where the spin down channel is strongly suppressed as well. Although the number of minority $3d$-states at the Fermi level is changing drastically among them, for all impurities it is significantly different to the Pt host LDOS~\cite{Suppl1}. Since the scattering is determined by the change of the electronic structure between impurity and host it leads to a suppression of the spin-down channel for all these impurities.
For Ni in Pt the situation is more complicated owing to the small exchange splitting
of the impurity $d$-states of Ni at $E_F$ and sensitivity of scattering to their exact position. This explains
the disagreement between the methods for Pt(Ni)~\cite{Suppl2}, otherwise rather convincing for the other cases. Based on
our analysis, we can formulate 
a necessary condition for an emergence of strong skew scattering
for {\it both} the AHE and the SHE in the same material: 
besides large effective SOC, 
there has to be a strong spin
asymmetry in the relative orbital composition of the host and impurity states at the Fermi energy. 

To glance at the microscopics of the skew-scattering process, we examine the distribution of the AHC
at the FS of two representative materials.  Namely, we compute the ``symmetrized" $k$-dependent AHC:
\begin{figure}[t!]
\includegraphics[width=0.99\linewidth]{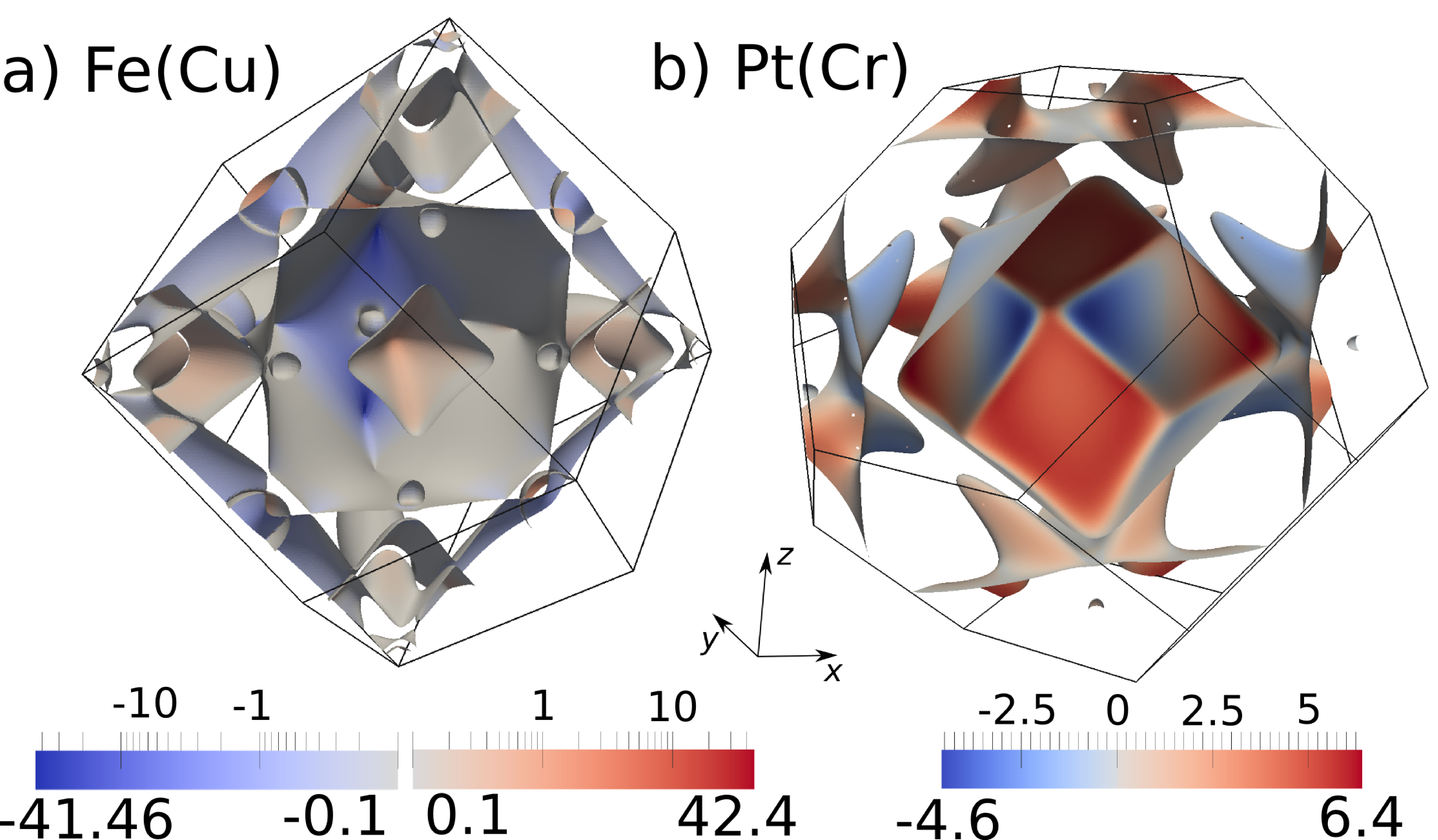}
\caption{
Fermi surface distribution of the symmetrized AHC, $\sigma^{\rm sym}_{yx}({\bf k})$ (in units of Bohr radius) in dilute alloys (a) Fe(Cu) and (b) Pt(Cr). Note the logarithmic color scale in (a). In both cases the scale spans the full range of values for the symmetrized integrand (Eq. \ref{Eq.:Asym}) on the Fermi surface.  \label{fig:Fermi}}
\end{figure}
\begin{equation}\label{Eq.:Asym}
  \sigma^{\rm sym}_{yx}({\bf k})= \sum_{\mu}{ \frac{ v_{y}({\bf k}) \, \lambda^{\mu}_{x}({\bf k}) + v_{y}({\bf k}') \, \lambda^{\mu}_{x}({\bf k}') }{ 2 \, \left\lvert {\bf v}({\bf k}) \right\rvert }  }~,
\end{equation}
where ${\bf k}=(k_1,k_2,k_3)$ and ${\bf k}'=(k_1,-k_2,k_3)$ are mirror images of each other 
with respect to the $y=0$ plane. 
For the nonmagnetic Pt host the sum is performed over the two 
degenerate
 bands ($\mu$=`$+$' or `$-$'), whereas this degeneracy is lifted for the magnetic Fe host and the sum can be omitted. Then the AHC can be obtained from Eq.~(\ref{Eq.:1})
where the integrand is replaced with $\sigma^{\rm sym}_{yx}({\bf k})$.
The symmetrized AHC captures the asymmetry between the scattering in $+y$ and $-y$ direction, and it would vanish identically without SOC. 

 The distribution of $\sigma^{\rm sym}_{yx}$ over the FS of Fe and Pt is shown in Fig.~\ref{fig:Fermi} for the Fe(Cu) and Pt(Cr) alloys. For Fe(Cu) the contributions to the AHE peak around small FS regions where the values of the symmetrized AHC are very large. Following Fabian and Das Sarma~\cite{Fabian1999}, we name such regions ``hot-spots". Here, the emergence of the hot-spots is due to the effect of
the weak SOC which is felt only at avoided crossings of the electronic bandstructure. On the other hand, in Pt(Cr) electrons experience skew scattering of opposite sign which is evenly
distributed over large parts of the FS  $-$ these are the 
so-called ``hot areas"~\cite{Zimmermann2012}. As opposed to Fe, here the effect
comes from strongly spin-orbit coupled spin-degenerate $d$-states at the Fermi energy. In a material like Pt(Cr), 
the hot areas, when integrated over the whole FS, can provide a gigantic contribution to the AHC.  In contrast, the singular
behaviour in a material like Fe(Cu) will be suppressed by vanishing area of the hot spot contributing to the integrated AHC. 
Generally, in complex transition-metals the two types of contributions can compete and the resulting 
values of the AHA can display a very non-trivial behaviour as a function of the Fermi level position, in analogy
to the intrinsic AHE~\cite{Nagaosa2010}. This can in turn lead to a large contribution of skew scattering to the anomalous
Nernst effect~\cite{Wimmer2014}.

In summary, we have shown that Boltzmann and Kubo-St\v{r}eda formalisms  agree in their 
description of skew scattering contributing to the AHE and SHE in the dilute limit of ferromagnetic alloys. We point
out that skew scattering is extremely sensitive to the fine details of the electronic structure which
motivates the use of {\it ab initio} schemes for studying its properties. By looking at the
chemical trends we study the interplay of the AHE and SHE in the same materials and formulate conditions for strong skew scattering in ferromagnetic alloys. These conditions are the strong effective spin-orbit coupling and the large spin asymmetry of the orbital character between impurity and host. Our work provides a 
necessary foundation for further material-specific studies of the skew-scattering in
ferromagnets, aimed at engineering the desired properties of spin-orbit driven transverse 
currents, which play a key role in modern spintronics.

This work was financially supported by the DFG projects	MO 1731/3-1 and SPP 1538 SpinCaT, and the HGF-YIG NG-513 project of the HGF. The J\"ulich group acknowledges computing time on the supercomputers at J\"ulich Supercomputing Center and 
JARA-HPC of RWTH Aachen University.
Furthermore, the work was partially supported by the Deutsche
Forschungsgemeinschaft (DFG) via SFB 762. In addition, M.G. acknowledges financial support from the DFG
via a research fellowship (GR3838/1-1) and the Leverhulme Trust via an Early Career Research Fellowship (ECF-2013-538). The Munich group acknowledges support by the DFG via SFB 689.

\bibliography{}

\end{document}